\documentclass[12pt]{iopart}

\usepackage{graphicx}
\usepackage{epsfig}
\usepackage{citesort}
\usepackage{xspace}
\usepackage{color}

\newcommand{\text}{\rm }
\newcommand{\gagg}{\ensuremath{g_{\text{a}\gamma}}\xspace}

\newcommand{\Rsun}{\ensuremath{R_\odot}\xspace}
\newcommand{\grtsim}{\,\rlap{\lower3.7pt\hbox{$\mathchar\sim$}}
\raise1pt\hbox{$>$}\,}
\newcommand{\lesssim}{\,\rlap{\lower3.7pt\hbox{$\mathchar\sim$}}
\raise1pt\hbox{$<$}\,}

\begin{document}

\title[An improved limit on the axion-photon coupling from the CAST
       experiment]
      {An improved limit on the axion-photon coupling from the 
       CAST experiment}

\newcommand{\CERN}{European Organization for Nuclear Research (CERN),
CH-1211 Gen\`eve 23, Switzerland}
\newcommand{\Saclay}{DAPNIA, Centre d'\'Etudes Nucl\'eaires de Saclay,
Gif-sur-Yvette, France}

\newcommand{\Darmstadt}{Technische Universit\"at Darmstadt,
    Institut f\"{u}r Kernphysik, Schlossgartenstrasse 9, 64289 Darmstadt,
  Germany}

\newcommand{\GSI}{Gesellschaft f\"ur Schwerionenforschung,
    GSI-Darmstadt, Plasmaphysik, Planckstr. 1, 64291 Darmstadt, Germany}

\newcommand{\MPE}{Max-Planck-Institut f\"{u}r extraterrestrische
  Physik, Geissenbachstrasse, 85748 Garching, Germany}

\newcommand{\Zaragoza}{Instituto de F\'{\i}sica Nuclear y Altas Energ\'{\i}as,
Universidad de Zaragoza, Zaragoza, Spain }
\newcommand{\Chicago}{Enrico Fermi Institute and KICP, University of Chicago,
Chicago, IL, USA}
\newcommand{\Thessaloniki}{Aristotle University of Thessaloniki, Thessaloniki,
Greece}
\newcommand{\Athens}{National Center for Scientific Research ``Demokritos'',
Athens, Greece}
\newcommand{\Freiburg}{Albert-Ludwigs-Universit\"{a}t Freiburg, Freiburg,
Germany}
\newcommand{\INR}{Institute for Nuclear Research, Russian Academy of
Sciences, Moscow, Russia}
\newcommand{\Vancouver}{Department of Physics and Astronomy, University of
British Columbia, Vancouver, Canada }
\newcommand{\Frankfurt}{Johann Wolfgang Goethe-Universit\"at, Institut f\"ur
Angewandte Physik, Frankfurt am Main, Germany}
\newcommand{\MPI}{Max-Planck-Institut f\"{u}r Physik
(Werner-Heisenberg-Institut), F\"ohringer Ring 6, 80805 M\"unchen, Germany}
\newcommand{\Zagreb}{Rudjer Bo\v{s}kovi\'{c} Institute,
Bijeni\v{c}ka cesta 54, P.O.Box 180, HR-10002 Zagreb, Croatia}
\newcommand{\Pisa}{Scuola Normale Superiore, Pisa, Italy}
\newcommand{\Kingston}{Department of Physics, Queen's University, Kingston,
            Ontario K7L 3N6, Canada}
\newcommand{\BNL}{Brookhaven National Laboratory, NY-USA}
\newcommand{\Sac}{DAPNIA, CEA-Saclay, Gif-sur-Yvette, France}
\newcommand{\Patras}{University of Patras, Patras, Greece}
\newcommand{\Lyon}{Inst. de Physique Nucl\'eaire, Lyon, France.}
\newcommand{\PAC}{Particle Astrophysics Center - Fermi National Accelerator
                  Laboratory, Batavia, IL~60510, USA}

\author{S~Andriamonje$^{2}$, S~Aune$^{2}$,
D~Autiero$^{1}$\footnote{Present address: \Lyon},
K~Barth$^{1}$, A~Belov$^{11}$, B~Beltr\'an$^{6}$\footnote{Present address:
\Kingston},
H~Br\"auninger$^{5}$, J~M~Carmona$^{6}$, S~Cebri\'an$^{6}$,
J~I~Collar$^{7}$, T~Dafni$^{2,4}$,
M~Davenport$^{1}$, L~Di~Lella$^{1}$\footnote{Present address: \Pisa},
C~Eleftheriadis$^{8}$, J~Englhauser$^{5}$\footnote{On leave},
G~Fanourakis$^{9}$, E~Ferrer~Ribas$^{2}$,
H~Fischer$^{10}$, J~Franz$^{10}$, P~Friedrich$^{5}$, T~Geralis$^{9}$,
I~Giomataris$^{2}$, S~Gninenko$^{11}$, H~G\'omez$^{6}$,
M~Hasinoff$^{12}$, F~H~Heinsius$^{10}$,
D~H~H~Hoffmann$^{3,4}$, I~G~Irastorza$^{2,6}$, J~Jacoby$^{13}$,
K~Jakov\v{c}i\'{c}$^{15}$, D~Kang$^{10}$, K~K\"onigsmann$^{10}$,
R~Kotthaus$^{14}$, M~Kr\v{c}mar$^{15}$, K~Kousouris$^{9}$,
M~Kuster$^{4,5}$, B~Laki\'{c}$^{15}$, C~Lasseur$^{1}$,
A~Liolios$^{8}$, A~Ljubi\v{c}i\'{c}$^{15}$, G~Lutz$^{14}$, G~Luz\'on$^{6}$,
D~Miller$^{7}$, A~Morales$^{6}$\footnote{Deceased}, J~Morales$^{6}$,
A~Ortiz$^{6}$,
T~Papaevangelou$^{1}$, A~Placci$^{1}$, G~Raffelt$^{14}$,
H~Riege$^{4}$, A~Rodr\'iguez$^{6}$, J~Ruz$^{6}$, I~Savvidis$^{8}$,
Y~Semertzidis$^{16}$\footnote{Permanent address: \BNL},
P~Serpico$^{17}$,
L~Stewart$^{1}$, J~Vieira$^{7}$, J~Villar$^{6}$, J~Vogel$^{10}$,
L~Walckiers$^{1}$
and K~Zioutas$^{1,16}$\\
(CAST Collaboration)}

\address{$^{1}$\CERN}
\address{$^{2}$\Saclay}
\address{$^{3}$\GSI}
\address{$^{4}$\Darmstadt}
\address{$^{5}$\MPE}
\address{$^{6}$\Zaragoza}
\address{$^{7}$\Chicago}
\address{$^{8}$\Thessaloniki}
\address{$^{9}$\Athens}
\address{$^{10}$\Freiburg}
\address{$^{11}$\INR}
\address{$^{12}$\Vancouver}
\address{$^{13}$\Frankfurt}
\address{$^{14}$\MPI}
\address{$^{15}$\Zagreb}
\address{$^{16}$\Patras}
\address{$^{17}$\PAC}

\ead{Milica.Krcmar@irb.hr}


\begin{abstract}
  We have searched for solar axions or similar particles that couple to two
  photons by using the CERN Axion Solar Telescope (CAST) setup with
  improved conditions in all detectors. From the absence of excess X-rays
  when the magnet was pointing to the Sun, we set an upper limit on the
  axion-photon coupling of 
  $g_{{\rm a}\gamma}<8.8\times 10^{-11}\,{\rm GeV}^{-1}$ 
  at 95\% CL for $m_{\rm a} \lesssim 0.02$~eV.
  This result is the best experimental limit over a broad range of
  axion masses and for $m_{\rm a} \lesssim 0.02$~eV also supersedes the
  previous limit derived from energy-loss arguments on globular-cluster
  stars.
\vspace{3mm}                                
\begin{flushleft} \textbf{Keywords}: axions, magnetic fields 
\end{flushleft}                             
\end{abstract}

\pacs{95.35.+d; 14.80.Mz; 07.85.Nc; 84.71.Ba}

\maketitle

\section{Introduction}                             \label{sec:intro}

The Peccei-Quinn (PQ) mechanism~\cite{Pec77,Pec77a,Peccei:2006as} 
remains perhaps
the most compelling explanation of the CP problem of strong interactions.
The existence of a new chiral $U(1)$ symmetry that is spontaneously broken
at some large energy scale $f_{\rm a}$ would allow for the dynamical
restoration of the CP symmetry in QCD. An inevitable consequence of this
mechanism is the existence of axions, the Nambu-Goldstone bosons of
$U(1)_{\rm PQ}$~\cite{Weinberg77,Wil78}. Axions are pseudoscalar particles with
properties closely related to those of neutral pions. 
In particular, their
mass and interaction strengths can be obtained by a simple rescaling from 
the corresponding $\pi^0$ properties (see equations~(\ref{eq3}) and
(\ref{eq:axionphoton}) below). 
Unsuccessful experimental
searches~\cite{Yao06} and astrophysical limits~\cite{Raffelt:2006cw} imply
that axions, if they exist, must be very light and very weakly interacting,
yet they are not necessarily harmless because they are a leading candidate
for the cold dark matter of the 
universe~\cite{Abb83,Pre83,Din83,Sikivie:2006ni}.
Moreover, several approaches to a unified description of dark energy and
dark matter involving pseudo Nambu-Goldstone bosons have been put forward
recently~\cite{Kim03,Mai04,Tak05}.

Experimental strategies to find ``invisible axions'' or more general
``axion-like particles'' (ALPs) include the cavity searches for galactic
dark matter axions~\cite{Sikivie:1983ip,Bradley:2003kg,Hag98,Asz04,Duf05},
searches for solar axions based on the ``helioscope''
method~\cite{Sikivie:1983ip,vanBibber:1988ge,Laz92,Mori98,Ino02,Zio05}, 
the Bragg
scattering technique~\cite{Paschos:1993yf,Avi98,Mor02,Ber01} or the resonant
method involving nuclear 
couplings~\cite{Krc98,Krc01,Lju04}, the polarization of light
propagating through a transverse magnetic
field~\cite{Maiani:1986md,Raffelt:1987im,Cam93,Zav05}, 
and the photon regeneration
method~\cite{Cam93,Ringwald:2003ns,Rabadan:2005dm,Kotz:2006bw,%
Lindner:2006,Cantatore:2006,Afanasev:2006cv,Afanasev:2006,%
Battesti:2006,Pugnat:2006,Sikivie:2007qm}.

Almost all of these past, present or future experiments rely on the axion
coupling to two photons that is described by the Lagrangian term
\begin{equation}
     {\cal L}_{{\rm a}\gamma}=
     -\frac{1}{4}\,g_{{\rm a}\gamma} F_{\mu\nu}\tilde F^{\mu\nu}a
     =g_{{\rm a}\gamma}\,{\bf E}\cdot{\bf B}\,a\,,
   \label{eq1}
\end{equation}
where $a$ is the axion field, $F$ the electromagnetic field-strength
tensor, $\tilde F$ its dual, ${\bf E}$ the electric, and ${\bf B}$ the
magnetic field. The axion-photon coupling strength is quantified by
\begin{equation}\label{eq2}
   g_{{\rm a}\gamma} =
  \frac{\alpha}{2 \pi}\, \frac{1}{f_{\rm a}}\,
  \left( \frac{E}{N} - \frac{2}{3}\,\frac{4+z+w}{1+z+w}\right) \,,
\end{equation}
where $z \equiv m_{\rm u}/m_{\rm d}$ and $ w \equiv m_{\rm u}/m_{\rm s} \ll
z$ are quark-mass ratios. The canonical values frequently used in the axion
literature are $z=0.56$ and $w=0.028$
~\cite{Gasser:1982ap,Leutwyler:1996qg}, although, for example, $z$ could
lie in the range 0.3--0.6~\cite{Yao06}. However, by far the largest
uncertainty in the relation between $g_{{\rm a}\gamma}$ and $f_{\rm a}$
comes from the model-dependent ratio $E/N$ of the electromagnetic and color
anomalies of the axial current associated with the axion field. Frequently
cited generic examples are the Kim-Shifman-Vainshtein-Zakharov (KSVZ)
model~\cite{Kim79,Shi80} where $E/N=0$ and the
Dine-Fischler-Srednicki-Zhitnitskii (DFSZ) model~\cite{Din81,Zhi80} where
$E/N=8/3$, but a much broader range of possibilities
exists~\cite{Cheng:1995fd}. The KSVZ model often serves as a prototype for
a hadronic model where axions do not couple directly to ordinary quarks and
charged leptons. 

The second parameter relevant for axion searches 
is the axion mass, that is related 
to the PQ symmetry breaking scale $f_{\rm a}$ by
\begin{equation}
   m_{\rm a} = \frac{z^{1/2}}{1+z}\, \frac{f_{\pi} m_{\pi}}{f_{\rm a}}
        =  6\,{\rm eV}\, \frac{ 10\,^6\,\rm GeV}{f_{\rm a}}\,,
   \label{eq3}
\end{equation}
where $z=0.56$ was assumed. Here, $m_{\pi} = 135$ MeV is the pion
mass and $f_{\pi} \approx 92$~MeV is its decay constant. In terms of
the axion mass, the axion-photon coupling is
\begin{equation}\label{eq:axionphoton}
 g_{{\rm a}\gamma}=\frac{\alpha}{2\pi}
 \left(\frac{E}{N}-\frac{2}{3}\,\frac{4+z}{1+z}\right)
 \frac{1+z}{z^{1/2}}\,\frac{m_{\rm a}}{m_\pi f_\pi} \,.
\end{equation}
This linear relationship between $g_{{\rm a}\gamma}$ and $m_{\rm a}$
defines the ``axion line'' for a given axion model in the 
$g_{{\rm a}\gamma}$--$m_{\rm a}$ parameter 
space\footnote{Depending on the value of $E/N$, the axion-photon coupling
can be positive or negative. In the following we always mean
$\left|g_{{\rm a}\gamma}\right|$ when we write $g_{{\rm a}\gamma}$, i.e.,
we always take $g_{{\rm a}\gamma}$ to be a positive number.}. 
Barring fine-tuned
cancellations, QCD axions are expected to lie near the narrow
band of parameters in the $g_{{\rm a}\gamma}$--$m_{\rm a}$ plane that is
defined by the corresponding $\pi^0$ properties.

Currently, laboratory searches for axions from the
Sun~\cite{Laz92,Mori98,Ino02,Avi98,Mor02,Ber01} 
are being extended by the CERN Axion
Solar Telescope (CAST). It takes advantage of the expected solar axion flux
that would be produced by the Primakoff effect, i.e., the conversion of
thermal photons in the Sun into axions in the presence of the electric
fields of the charged particles in the solar interior. CAST itself then
uses the reverse process of coherent axion-photon conversion in a
``magnetic telescope,'' i.e., in the bores of a 9.26~m long LHC dipole
magnet oriented toward the Sun. CAST has recently reported the best
experimental limit obtained so far of $g_{{\rm a}\gamma} < 1.16 \times
10^{-10} \; {\rm GeV}^{-1}$ for $m_{\rm a} \lesssim0.02$~eV~\cite{Zio05}.
This limit is comparable to the most severe constraint from the population
of horizontal-branch (HB) stars in globular clusters that implies 
$g_{{\rm a}\gamma} \lesssim 10^{-10} \; {\rm GeV}^{-1}$
~\cite{Raffelt:2006cw,Raf96}.

Here we report new results from the CAST experiment obtained in 2004 with
improved conditions in all detectors. In addition, the solar axion flux is
calculated from a modern solar model for the purpose of the CAST data
interpretation. For the first time we obtained better sensitivity than that
arising from energy-loss arguments on globular cluster stars.

The result reported here provides our final limit for the CAST vacuum setup
(so-called CAST phase~I)
where the sensitivity is essentially limited to $m_{\rm a}\lesssim0.02$~eV
as explained below. This implies that CAST has not yet reached the ``axion
line'' in the $g_{{\rm a}\gamma}$--$m_{\rm a}$ space, but rather provides
the most restrictive limits on the two-photon vertex of axion-like
particles that are somewhat lighter for a given interaction strength that
expected for QCD axions.  Meanwhile, CAST has been refurbished as a tuning
experiment, planning to explore the mass region up to about 
1~eV (CAST phase II). CAST will then probe QCD axion models
and will also be sensitive to the existence of large extra
dimensions~\cite{Hor04}, introduced in the world-brane scenarios to solve
the hierarchy problem in particle physics~\cite{Ark98}.

We begin in section~\ref{sec:axflux} with a new calculation of the solar
axion flux. In section~\ref{sec:CAST} we briefly describe the CAST
experiment. In section~\ref{sec:data} we report the 2004 measurements and
their analysis. 
In section~\ref{sec:CASTphaseI} we present the combined
result of the CAST phase~I and conclude in section~\ref{sec:conclusion}.

\section{Solar axion flux}                          \label{sec:axflux}

\subsection{General expression}
\label{sec:general-expression} 
Particles that interact with photons by the Lagrangian of
equation~(\ref{eq1}) are produced in a hot thermal plasma by virtue of the
fluctuating electric and magnetic fields, i.e., the fluctuations of ${\bf
  E}\cdot{\bf B}$ in the plasma act as a source for the production of
axions or similar particles. Scalar particles couple to $({\bf E}^2+{\bf
  B}^2)$ instead of ${\bf E} \cdot {\bf B}$ for pseudoscalars. Hence for
scalars, the fluctuations of ${\bf E}^2$ and ${\bf B}^2$ would play the
analogous role. In principle, the axion emission rate for a given
temperature, density and chemical composition can be calculated directly
from the spectrum of ${\bf E}\cdot{\bf B}$ fluctuations as was done for a
degenerate plasma in~\cite{Altherr:1993zd}. For the purpose of calculating
the solar axion flux, this formalism gives the same result as the simpler
approach where one calculates different ``processes'' contributing to the
emission rate~\cite{Raffelt:1985nk}. The dominant contribution derives from
those fluctuations where ${\bf E}$ is provided by the charged particles of
the medium whereas ${\bf B}$ comes from propagating thermal photons. This
is the usual Primakoff process where a photon converts into an axion in the
electric field of a charged particle. Of course, in a plasma the charged
particles are correlated in such a way that charges are screened beyond a
certain distance, an effect that 
is crucial to take into account. 
Other emission processes that are subdominant include the
``electro-Primakoff effect'' where the ${\bf B}$ field is provided by a
moving electric charge, i.e., the axion is produced from two virtual
photons, each attached to a charged particle~\cite{Raffelt:1985nk}. In the
Sun, all particles, including the electrons, are nonrelativistic so that
the ${\bf B}$ fields associated with moving electric charges are small,
explaining that the electro-Primakoff effect is much less important than
the ordinary Primakoff effect involving real photons.

Axions or similar particles typically will also interact with other
particles in the plasma, notably the electrons and nuclei. However, in the
absence of a signal it is conservative to use only those processes at the
source that are implied by the very interaction structure that is used for
the detection, i.e., the axion-two photon vertex. Moreover, for QCD 
axion models, the solar fluxes derived from other processes are expected to
be much smaller than the Primakoff-induced flux if we take account of
stellar energy-loss limits on these interactions. Therefore, our results
are especially relevant for hadronic axions that do not couple to electrons
at tree level so that the Primakoff process is expected to be the dominant
process and a non-observation of axions by CAST provides new and nontrivial
limits.

The transition rate for a photon of energy $E$ into an axion of the same
energy by the Primakoff effect in a stellar plasma is~\cite{Raffelt:1987np}
\begin{equation}\label{eq:Gamma-ag}
  \Gamma_{\gamma\to{\rm a}}=
  \frac{g_{{\rm a}\gamma}^2T\kappa_{\rm s}^2}{32 \pi}
  \bigg[\bigg(1+\frac{\kappa_{\rm s}^2}{4E^2}\bigg)
  \ln\bigg(1+\frac{4E^2}{\kappa_{\rm s}^2}\bigg)-1\bigg]\,,
\end{equation}
where $T$ is the temperature (natural units with $\hbar=c=k_{\rm B}=1$ are
used). Recoil effects are neglected so that the photon and axion energies
are taken to be equal.  The screening scale in the Debye-H\"uckel
approximation~is
\begin{equation}
  \kappa_{\rm s}^2=
  \frac{4\pi\alpha}{T}\biggl(n_{\rm e}+
  \sum_{\rm nuclei}Z_j^2n_j\biggr),
\end{equation}
where $n_{\rm e}$ is the electron number density and $n_j$ the number
density of the $j$-th ion of charge $Z_j$.

The total axion number flux at the Earth (average solar distance $D_\odot$ from
the Earth)~is
\begin{equation}
  \Phi_{\rm a}=\frac{R_\odot^3}{4\pi D_{\odot}^2} \int_0^1
  \!\!{\rm d}r\,4\pi\,r^2\int_{\omega_{\rm pl}}^\infty \!\!{\rm d}E\,
  \frac{4 \pi k^2}{(2\pi)^3}\,
  \frac{{\rm d}k}{{\rm d}E}\,2f_{\rm B}\, \Gamma_{\gamma\to{\rm
  a}}\;,
\end{equation}
where $f_{\rm B}= ({\rm e}^{E/T}-1)^{-1}$ is the Bose-Einstein distribution
of the thermal photon bath in the solar plasma and $r=R/R_\odot$ is a
dimensionless solar radial variable, normalized to the solar radius
$R_{\odot}$. We take into account the plasma frequency for the photons in
the system
\begin{equation}
  \omega_{\rm pl}^2=\frac{4 \pi \alpha n_{\rm e}}{m_{\rm e}}\,.
\end{equation}
It enters the dispersion relation as $E^2=k^2+\omega_{\rm pl}^2$ so
that ${\rm d}k/{\rm d}E=E/k$. Of course, the value of the plasma
frequency depends on the radial position in the Sun.

The approximations used in this calculation give us the axion flux for the
range of X-ray energies relevant for CAST and also give us the total number
flux and total energy loss rate, with an estimated precision of a few
percent. However, at energies near and below a typical solar plasma
frequency, i.e., for energies near or below 0.3~keV, this calculation is
not appropriate because the charged particles were treated as static
sources of electric fields, neglecting both recoil effects and collective
motions. The charged particles of the plasma themselves undergo plasma
oscillations with frequencies near $\omega_{\rm pl}$. Therefore, even
though the thermal photon energies are cut off below $\omega_{\rm pl}$, the
emitted axion energies can reach down to zero~\cite{Raffelt:1987np},
although with a steeply falling spectrum. For applications where this
low-energy flux might be of interest, perhaps the small solar axion flux at
optical energies, a new calculation is needed that properly includes the
dynamical aspects of the collective plasma motions, not only the static
screening effects that are included here.

The X-ray telescope used in CAST is an imaging device so that we do not
necessarily want to treat the Sun as a point source. We determine the
differential solar axion flux as an apparent surface luminosity
$\varphi_{\rm a}(E,r)$ of the solar disk.  Put another way, we determine
the flux per unit surface area of the two-dimensional solar disk as a
function of the dimensionless radial coordinate $0\leq r\leq1$. The total
axion number flux at the Earth~is
\begin{equation}\label{eq:fluxintegration}
  \Phi_{\rm a}=2\pi \int_0^1 {\rm d}r\,r\,
  \int_{\omega_{\rm pl}}^\infty {\rm d}E\,\varphi_{\rm a}(E,r)\,,
\end{equation}
i.e.,\ $\varphi_{\rm a}$ is in units of $\rm cm^{-2}~s^{-1}~keV^{-1}$ per
unit surface area.  The unit surface area is dimensionless because the
radial coordinate $r$ on the solar disk is dimensionless. 
Of course, instead of using a dimensionless variable $r$ that is normalized to
the solar radius, we could use one that is normalized to the apparent
angular radius of the Sun in the sky. In this case, $\varphi_{\rm a}$ is to
be multiplied with $(R_\odot/D_\odot)^2$, giving us the surface flux per
steradians of solid angle 
and therefore the ${\rm d}r$-integration 
in equation~(\ref{eq:fluxintegration}) would extend to
the angular radius $R_\odot/D_\odot$ of the Sun.

To determine the surface luminosity $\varphi_{\rm a}$ we re-write the
integral over the solar volume as
\begin{equation}
  4\pi \int_0^1 {\rm d}r\,r^2 \to 2\pi\int_0^1 {\rm d}r\,r
  \int_{{\rm line~of}\atop{\rm sight}} {\rm d}z\,.
\end{equation}
We assume the Sun to be projected on a 2-dimensional disk, i.e., we neglect
parallax effects, where the ratio $R_{\odot}/D_{\odot}\approx 0.5\%$ sets
the accuracy level of this approximation. We may then write the volume
integral as
\begin{equation}
   \int_{{\rm line~of}\atop{\rm sight}} {\rm d}z
  =2\int_{r}^{1}{\rm d}\rho\,\frac{\rho}{\sqrt{\rho^2-r^2}}\,,
\end{equation}
where $\rho$ is the dimensionless radial position in the Sun where the
physical variables (temperature, density, etc.) have to be taken from a
solar model. Altogether we find
\begin{equation}
  \label{eq:axion-surface-luminosity}
  \varphi_{\rm a}(r,E)=\frac{R_\odot^3}{2\pi^3D_{\odot}^2} \int_r^1
  {\rm d}\rho\,\frac{\rho}{\sqrt{\rho^2-r^2}}\,
  E\,k\,f_{\rm B}\,\Gamma_{{\rm a}\to\gamma}\,,
\end{equation}
where $f_{\rm B}$ and $\Gamma_{{\rm a}\to\gamma}$ are to be taken at
the position $\rho$.

\subsection{Integration over solar models}
\label{sec:integration-over-solar}
In order to perform these integrations we use a recent solar model for
which detailed data are publicly available, the 2004 model of Bahcall and
Pinsonneault~\cite{Bahcall:2004fg}, which is tabulated on a fine grid (1071
points). The axion flux parameters~are found to~be
\begin{eqnarray}
  \Phi_{\rm a}&=&g_{10}^2\,
  3.75\times10^{11}~{\rm cm}^{-2}~{\rm s}^{-1}\,,
  \nonumber\\
  L_{\rm a}&=&g_{10}^2\,1.85\times 10^{-3} L_\odot\,,\nonumber\\
  \langle E\rangle&=&
  4.20~{\rm keV}\,,\nonumber\\
  \langle E^2\rangle&=&
  22.7~{\rm keV}^2\,,
\end{eqnarray}
where $g_{10}=g_{{\rm a}\gamma}/10^{-10}~{\rm GeV}^{-1}$. The maximum of
the distribution is at $3.00$~keV. For self-consistency we have used the
values of the solar radius ($R_{\odot}=6.9598\times 10^{10}~{\rm cm}$) and
of the solar luminosity 
($L_{\odot}=3.8418\times 10^{33}~{\rm erg}~{\rm s}^{-1}$)
from~\cite{Bahcall:2004fg}.
\begin{figure}[t]
    \centerline{\includegraphics[width=0.5\textwidth]{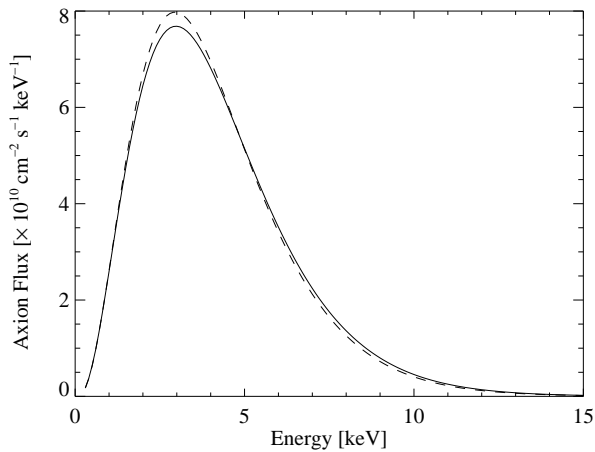}}
    \caption{Comparison of the solar axion flux calculated from a
      modern solar model~ \cite{Bahcall:2004fg} ($\full$) and an older
      solar model published in 1982~\cite{Bahcall:1981zh} ($\broken)$.
      An axion-photon coupling of $1\times 10^{-10}$~GeV$^{-1}$
       is assumed.}  
  \label{fig:solar-flux}
\end{figure}

In figure~\ref{fig:solar-flux} we compare the differential axion flux
obtained from this model with that from the 1982
model~\cite{Bahcall:1981zh} that was used in an earlier calculation of the
solar axion flux~\cite{vanBibber:1988ge}. Even though a large number of
details has changed, standard solar models and their neutrino predictions
have remained remarkably robust.  The same is true for the axion flux
prediction that depends only mildly on the exact solar model. Even though
details of our understanding of the Sun may change in the future, the solar
neutrino measurement of all active flavors~\cite{Ahmed:2003kj} has provided
a direct verification of the Sun's inner properties so that it is hard to
imagine that adopting a standard solar model for the axion flux calculation
could be a source of gross error.

An analytic approximation to the solar axion flux spectrum may sometimes be
useful. Instead of the analytic form proposed by van Bibber et
al.~\cite{vanBibber:1988ge}, we find that an excellent fit is provided by a
three-parameter function of the kind
\begin{equation}\label{fitchoice}
  \frac{{\rm d}\Phi_{\rm a}}{{\rm d}E}=C\,\bigg(\frac{E}{E_0}\bigg)^{\beta}
  {\rm e}^{-(\beta+1)E/E_0}\,.
\end{equation}
Here, $C$ is a normalization constant while the fit parameter $E_0$
  coincides with the average energy, $\langle E\rangle=E_0$.  If we match
  the number flux, average energy, and width of the numerical spectrum with
  such a fit, we find (energies in keV)
\begin{equation}\label{bestfit}
  \frac{{\rm d}\Phi_{\rm a}}{{\rm d}E}=6.02\times
  10^{10}~{\rm cm}^{-2}~{\rm s}^{-1}~{\rm keV}^{-1}\, g_{10}^2
  \,E^{2.481}{\rm e}^{-E/1.205}\,.
\end{equation}
The fit is accurate at better than about the $1\%$ level in the interval 1
to 11~keV, while the relative accuracy is bad at very low or very high
energies.

\begin{figure*}
  \begin{minipage}{0.49\textwidth}
      \centerline{\includegraphics[width=0.99\textwidth]
                                  {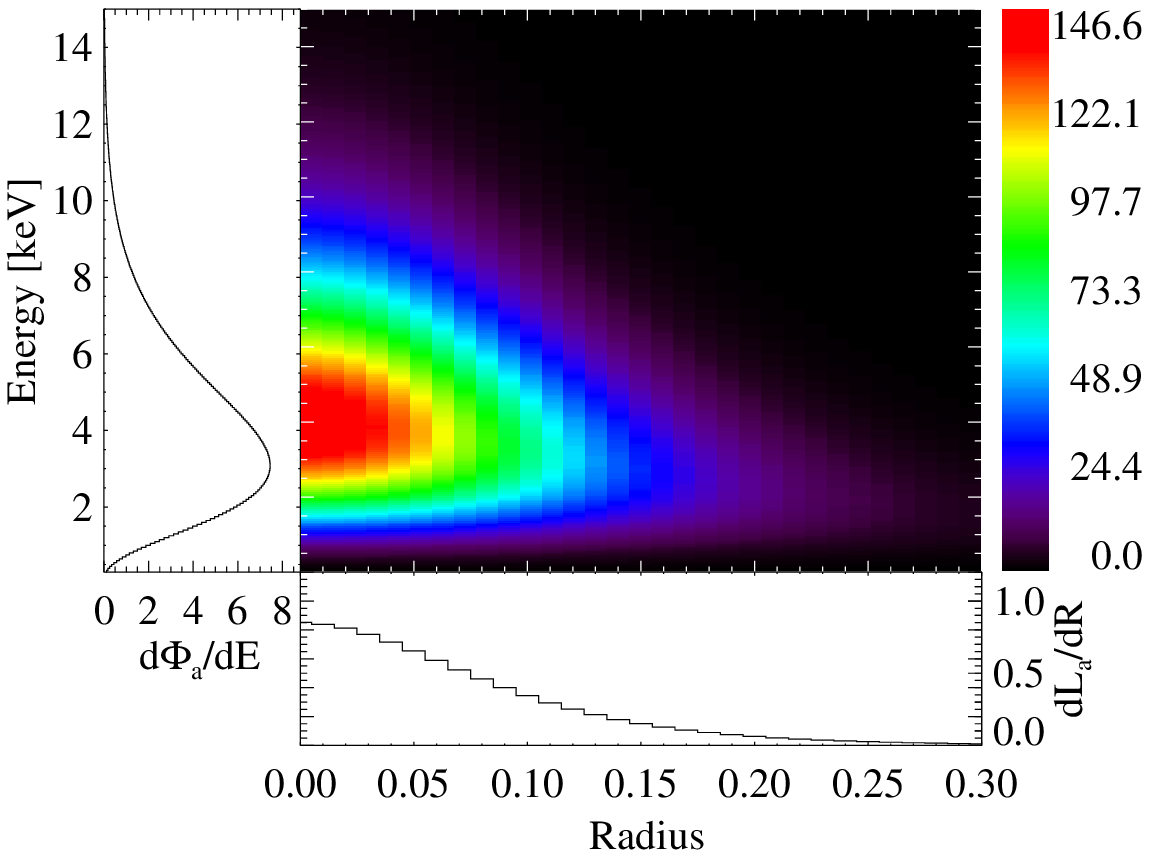}}
  \end{minipage}
  \begin{minipage}{0.49\textwidth}
      \centerline{\includegraphics[width=0.99\textwidth]
                                  {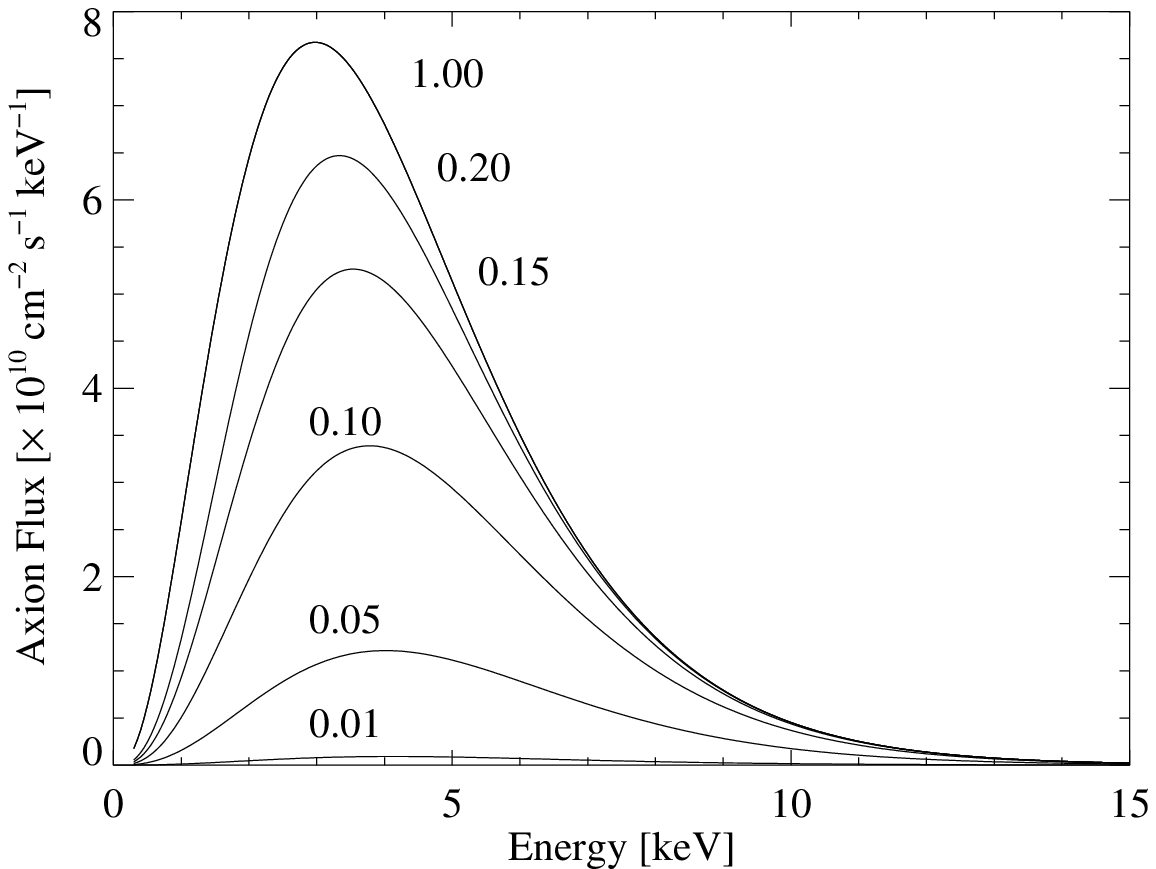}}
  \end{minipage}
  \caption{\label{fig:axionspec-radius-dependance} Left: Solar axion surface
    luminosity depending on energy and the 
    radius $r$ on the solar disk. The flux is given in units of
    $\text{axions}\,\text{cm}^{-2}\,\text{s}^{-1}\,\text{keV}^{-1}$
    per unit surface area on the solar disk.
    Also shown is the radial distribution of
    the axion energy loss rate of the Sun (${\rm d}L_{\text{a}}/{\rm d}R$)
    as well as the energy
    distribution of the solar axion flux (${\rm d}\Phi_{\text{a}}/{\rm d}E$). 
    Right: Differential solar axion
    spectrum, derived by integrating the model shown on the left up to
    different values of $r$ in units of the solar radius $R_\odot$. 
    The peak of the spectrum moves towards lower
    energies if integration radius moves towards the outer rim of the solar
    disk.}
\end{figure*}

In figure~\ref{fig:axionspec-radius-dependance} we show a contour plot of
the solar surface luminosity in axions $\varphi_{\rm a}(E,r)$ as a function
of axion energy $E$ (keV) and a dimensionless radial coordinate $r$ on the
solar disk.  
We also show the solar axion flux as a function of the energy for
several values of $r$ 
for an axion-photon coupling $\gagg =
1\times10^{-10}\,\text{GeV}^{-1}$.  Most of the axion flux emerges from the
inner 20\% of the solar disk.

\subsection{Do axions escape from the Sun?}
\label{sec:do-axions-escape-from-the-sun}
CAST can detect solar axions only if these particles actually escape
from the Sun after production. To estimate the axion mean free path
(mfp) in the Sun we note that the photon-axion conversion rate of
equation~(\ref{eq:Gamma-ag}) is identical (in natural units) with
the inverse mfp of a photon with energy $E$  
based on the Primakoff process. 
Since we work in the recoil-free approximation where the
energy of the photon and axion are identical,
equation~(\ref{eq:Gamma-ag}) also gives us the inverse mfp for the
reverse process of an axion with energy $E$ to be converted to a
photon, i.e., of axion absorption. As an example we consider an
axion with energy 4~keV, near the average of the expected spectrum,
and note that the temperature at the solar center is $T\approx
1.3$~keV whereas the screening scale is $\kappa_{\rm
s}\approx9$~keV. The axion mfp is then found to be $\lambda_{\rm
a}\approx g_{10}^{-2}\,6\times10^{24}~{\rm cm}\approx
g_{10}^{-2}\,8\times10^{13}\,R_{\odot}$, or about $10^{-3}$ of the
radius of the visible universe.

Therefore, in the absence of other interactions, the axion-photon
coupling would have to be more than $10^7$ times larger than the
CAST limit for axions to be re-absorbed within the Sun. However,
even in this extreme case axions are not harmless for the solar
structure because they would then carry the bulk of the energy flux
within the Sun that otherwise is carried by photons. Low-mass
particles that are trapped in the Sun should interact so strongly
that their mfp is smaller than that of
photons~\cite{Raffelt:1988rx}. Note that the photon mfp for the
conditions near the solar center is less than 1~mm. Only particles
with a mfp not much larger than this will leave the solar structure
unaffected. They will cause a gross acceleration of the rate of
energy transfer in the Sun and other stars if their mfp is much
larger than this. These requirements are so extreme that for
anything like axions the possibility of re-absorption is not a
serious possibility. 

\subsection{Is the CAST limit consistent with standard solar models?}
\label{sec:is-the-cast-limit-consistent} 
Our final limit, equation~(\ref{finalresult}), on the axion-photon coupling 
implies that the solar axion flux is bounded by $L_{\rm a}
\lesssim1.3\times10^{-3}\,L_\odot$.  Self-consistent solar models
including axion losses were constructed in~\cite{Schlattl:1998fz}.
In particular, it was found that helioseismological measurements of
the sound-speed profile do not exclude $g_{{\rm a}\gamma}$ much
below $1\times10^{-9}~{\rm GeV}^{-1}$ where $L_{\rm a}$ would be
around $0.2\,L_\odot$. It was also found that for $g_{{\rm
a}\gamma}=4.5\times10^{-10}~{\rm GeV}^{-1}$, where $L_{\rm
a}=0.04\,L_\odot$, the $^8$B solar neutrino flux would exceed the
standard prediction by about 20\%. While the all-flavor $^8$B flux
has been measured by SNO with a precision of about 8.8\%
\cite{Ahmed:2003kj}, the uncertainties of the flux prediction are of
order 20\% \cite{Bahcall:2004pz}. We conclude that the solar axion
flux corresponding to the CAST limit is far below the range where it
would affect helioseismology or the measured solar neutrino flux in
significant ways. For the purpose of CAST it is therefore justified
to treat the solar axion losses as a negligible perturbation of the
standard solar models.

\section{CAST experiment}                         \label{sec:CAST}

A detailed description of a proposed method of detecting solar
axions by the CERN Axion Solar Telescope has been given
elsewhere~\cite{Zio05,Zio99}. Here we only recall that CAST uses a
decommissioned Large Hadron Collider (LHC) prototype magnet with a
field of $B = 9.0$ T in the interior of two parallel pipes of length
$L = 9.26$ m and a cross-sectional area $A = 2\times 14.5 \; {\rm
cm}^2$. The magnet is mounted on a moving platform with
low-background X-ray detectors on either end, allowing it to track
the Sun about 3 hours per day. Solar axions of a broad energy
spectrum which peaks at $E\sim4$~keV, essentially reflecting the
thermal conditions in the solar interior, may be converted into real photons
inside the transverse magnetic field. The probability of this
conversion in vacuum~\cite{vanBibber:1988ge} is
\begin{equation}
      P_{{\rm a}\to\gamma}=\left( \frac{g_{{\rm a}\gamma}B}
          {q} \right)^2 \sin^2\left( \frac{qL}{2} \right)\,,
    \label{eq4}
\end{equation}
where $q =m_{\rm a}^2/2E$ is the photon-axion momentum difference.
Therefore, the axion signal should appear as an excess of photons above
background in three different X-ray detectors, a conventional time
projection chamber (TPC), a gaseous chamber micromegas (MM), and an X-ray
telescope with a charge coupled device (CCD) as focal plane detector, when
the Sun and magnet are aligned.  A differential flux of photons of
\begin{eqnarray}\label{eq5}
 \frac{{\rm d}\Phi_{\gamma}}{{\rm d}E}&=&
 \frac{{\rm d}\Phi_{\rm a}}{{\rm d}E}\,P_{{\rm a}\to\gamma}
 \\
 &=& 0.088~{\rm cm^{-2}~day^{-1}~keV^{-1}}\,
 g_{10}^4\, E^{\,2.481}{\rm e}^{-E/1.205}
 \left(\frac{L}{9.26~\rm m}\right)^2
 \left(\frac{B}{9.0~\rm T}\right)^2\nonumber
\end{eqnarray}
is expected at the end of the magnet in case of the coherent
conversion which occurs for $qL \lesssim 1$, providing for the CAST
sensitivity to the axion mass of up to 0.02~eV.  Here ${\rm
d}\Phi_{\rm a}/{\rm d}E$ is the axion spectrum at the Earth,
expressed by the analytic approximation~(\ref{bestfit}). For
$qL\grtsim 1$ the sensitivity is reduced owing to the photon-axion
momentum mismatch. To search for more massive axions, coherence can
be restored by filling the magnetic conversion region with a low-$Z$
gas whose pressure is adjusted in such a way that plasma frequency,
which acts as an effective mass for the photon, equals the axion
mass~\cite{vanBibber:1988ge}. As one can see from
equation~(\ref{eq5}), due to the $g_{{\rm a}\gamma}^4$ rate
suppression, a lot of effort is required to improve the existing
limits on axions significantly.

\section{Measurement and analysis}      \label{sec:data}

The new data taking has been performed after dismounting and
remounting the entire setup for the X-ray detection, with some
changes in the detectors, in the materials placed nearby, and in the
operating conditions with respect to the data taking considered in
our previous paper~\cite{Zio05}.

\subsection{X-ray telescope and CCD detector}
\label{sub:CCD}

The X-ray mirror telescope with a CCD as the focal plane
detector is placed behind one of the magnet bores at the 
west end of the magnet and is looking for X-rays from ``sunrise'' axions. 
The sensitivity of the X-ray telescope is significantly
improved compared to the 2003 data taking period by a continuous monitoring
of its pointing stability, which allows us to exploit the full potential of
the system. As a consequence, the area on the CCD where the axion signal is
expected could be reduced by a factor of 5.8 compared to 2003. In
addition, after adding new internal and external shielding components, the
background level has been reduced by a factor of 1.5 in comparison with the
2003 setup. The CCD detector has shown a stable performance over the entire
2004 running period and has acquired 197~h of tracking data (when the
magnet was pointing to the Sun) and 1890~h of background data (taken from
the same area during the non-tracking periods).  For a detailed description
of the X-ray telescope design, its performance, and background systematics
during the 2004 data taking period we refer to~\cite{Kus06}. Although no
significant excess signal over background could be detected with the X-ray
telescope during the data taking period of 2004, a new upper limit on the
axion-photon coupling could be derived. In the following sections we
summarize the analysis techniques and results we obtained for the X-ray
telescope.

\begin{figure*}[!ht]
  \centering
  \begin{minipage}{0.49\textwidth}
    \centerline{\includegraphics[height=0.6\textheight,angle=0]
      {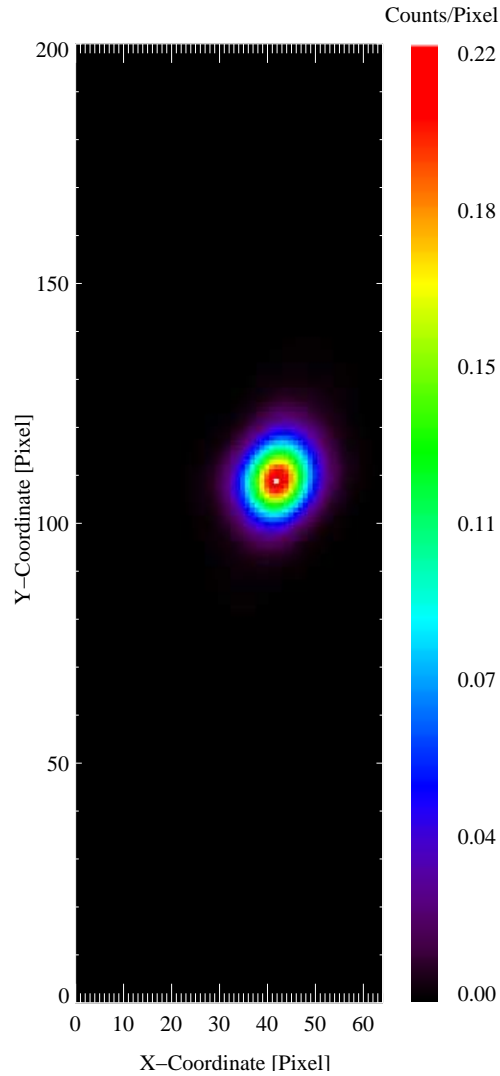}}
  \end{minipage}
  \caption{\label{fig:ccd-backrgound-distribution-axion-spot} 
    Expected ``axion'' image of the Sun as it would be observed by the CCD
    detector assuming the axion surface luminosity shown in
    figure~\ref{fig:axionspec-radius-dependance} and assuming zero detector
    background. To determine the axion spot on the CCD, the PSF of the
    mirror system and the total effective area of the X-ray telescope were
    taken into account. The count rate integrated over the region of the
    spot is normalized to unity.}
\end{figure*}

\subsubsection{Expected axion signal.}
\label{ssub:axion}
As described in section~\ref{sec:integration-over-solar} the solar
axion surface luminosity $\varphi_{\rm a}(E,r)$ is a function of
axion energy $E$ and the distance from the center of the solar disk
$r$, see figure~\ref{fig:axionspec-radius-dependance} and
equation~(\ref{eq:axion-surface-luminosity}). The differential axion
spectrum emitted by a circular region on the solar disk can be
calculated by integrating $\varphi_{\rm a}(E,r)$ from the center of
the solar disk up to a radius $r$, resulting in spectra as shown
in the right part of figure~\ref{fig:axionspec-radius-dependance}.
It is apparent that not only the integral flux changes, but the peak
energy of the spectrum slightly shifts towards lower energies for
increasing $r$. For $r=1$ the spectrum peaks at lowest
energies. Therefore, a photon signal resulting from axion-to-photon
conversion in the magnet bore subsequently observed by the CCD would
show a similar energy and radial intensity dependence, after being
modified by the spatial response function of the X-ray optics (point
spread function -- PSF) and the energy response of the CCD detector.

To characterize the spatial intensity distribution of a potential axion
signal on the CCD detector, we multiplied the axion solar surface
luminosity by $P_{{\rm a}\to\gamma}$ of equation~(\ref{eq4}) and folded the
resulting distribution with the point spread function of the X-ray
telescope. 
The corresponding intensity distribution (axion spot image) 
we would expect on the CCD chip is shown in 
figure~\ref{fig:ccd-backrgound-distribution-axion-spot}. 
The axion spot image is slightly distorted from a mere
circular shape due to the PSF of the X-ray mirror optics~\cite{Kus06}. In
addition, the image is enlarged compared to an axion solar spot image
multiplied with the scale factor of the mirror optics assuming a perfect
linear imaging characteristics. The latter effect is a direct consequence
of the finite spatial resolution of the optics, the spatial resolution of
the detector (pixel size), and the limited pointing accuracy of the CAST
tracking system ($\lesssim 0.01^{\circ}$) 
which spread the image of a point source. For our simulations 
we did not take into account either that the PSF depends on
photon energy or the fact that it is in general a function of the position
on the detector (off-axis angle). We verified that both effects are of
minor importance for CAST, because the image is close to the optical axis,
i.e., we have to deal with small off-axis angles, and we expect a signal in
a relative narrow energy band ($1$--$7$\,\text{keV}).

For a fixed $m_{\rm a}$, the measured count-rate spectrum from
axion-to-photon conversion is calculated by
\begin{equation}
  \label{eq:count-spectrum}
  s_{i}=\int_{E_i}^{E_{i+1}}\,{\rm d}E'\int_0^\infty
  A(E)\,R(E',E)\;2\pi
  \int_0^{r_{\rm s}}
  \frac{{\rm d}\Phi_{\gamma}(E,\gagg,r)}{{\rm d}E}\,
  r\, {\rm d}r\,{\rm d}E\;,
\end{equation}
where $s_i$ is the count rate detected in detector energy channel $i$ in
units of counts~s$^{-1}$, $A(E)$ is the effective area of the X-ray
telescope in units of $\text{cm}^2$, $R(E,E')$ is the detector response
function, and ${\rm d}\Phi_{\gamma}(E,\gagg,r)/{\rm d}E$ is the
differential spectrum of photons resulting from axion-to-photon conversion
in the magnet for a given $m_{\text{a}}$, $g_{\text{a}\gamma}$, and the
radius $r$ in units of the solar radius $\Rsun$.  The radius $r$ directly
corresponds to the radius of a circular signal area $r_{\text{s}}$ we used
to extract the potential signal, 
where $r_{\text{s}}$ is given in detector
pixel coordinates.  This signal area is centered at the location where we
expect the maximum signal intensity on the CCD chip. The effective area of
the X-ray telescope includes the quantum efficiency of the detector, the
mirror reflectivity, the influence of the finite size of the axion emission
region on the Sun, and geometric effects due to, e.g., the finite diameter
of the magnet bore (e.g. vignetting, see~\cite{Kus06}).

\begin{figure*}[t]
  \begin{minipage}{0.49\textwidth}
    \centerline{\includegraphics[width=0.95\columnwidth]{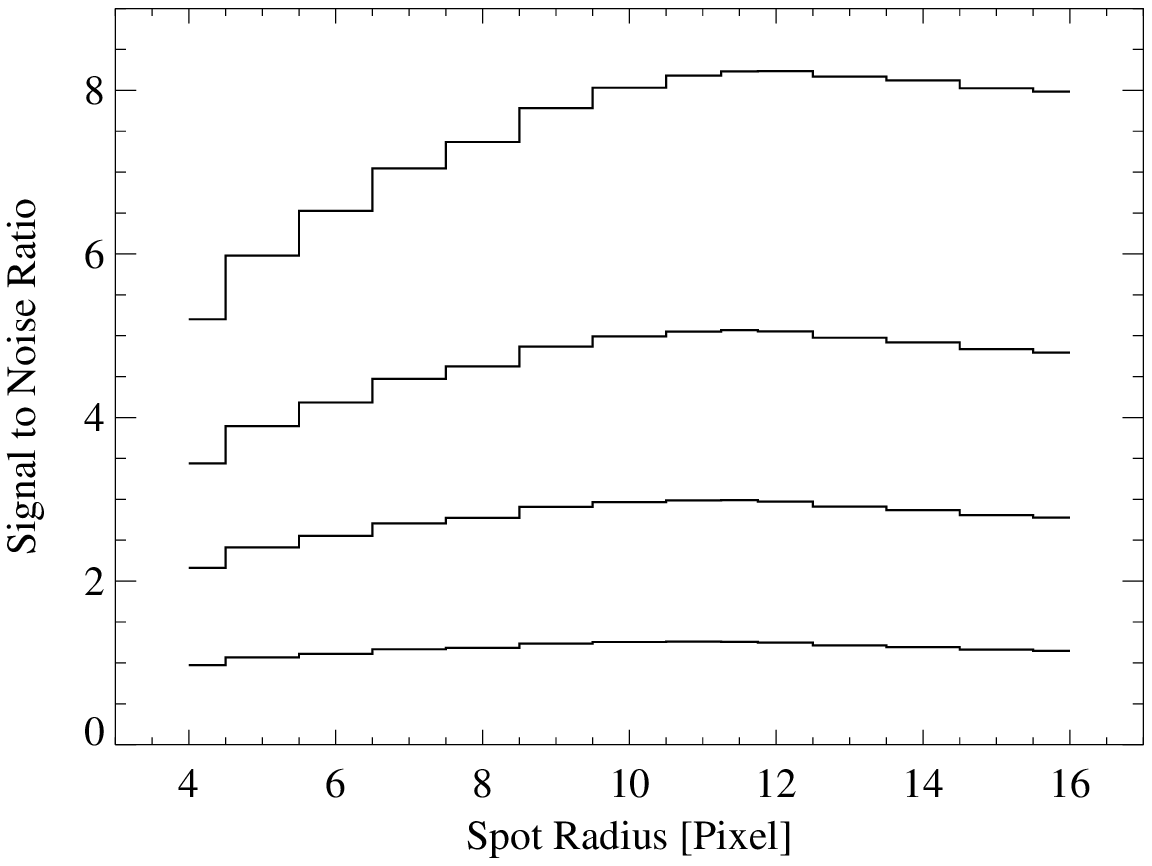}}
  \end{minipage}
  \begin{minipage}{0.49\textwidth}
    \centerline{\includegraphics[width=0.95\textwidth]
                 {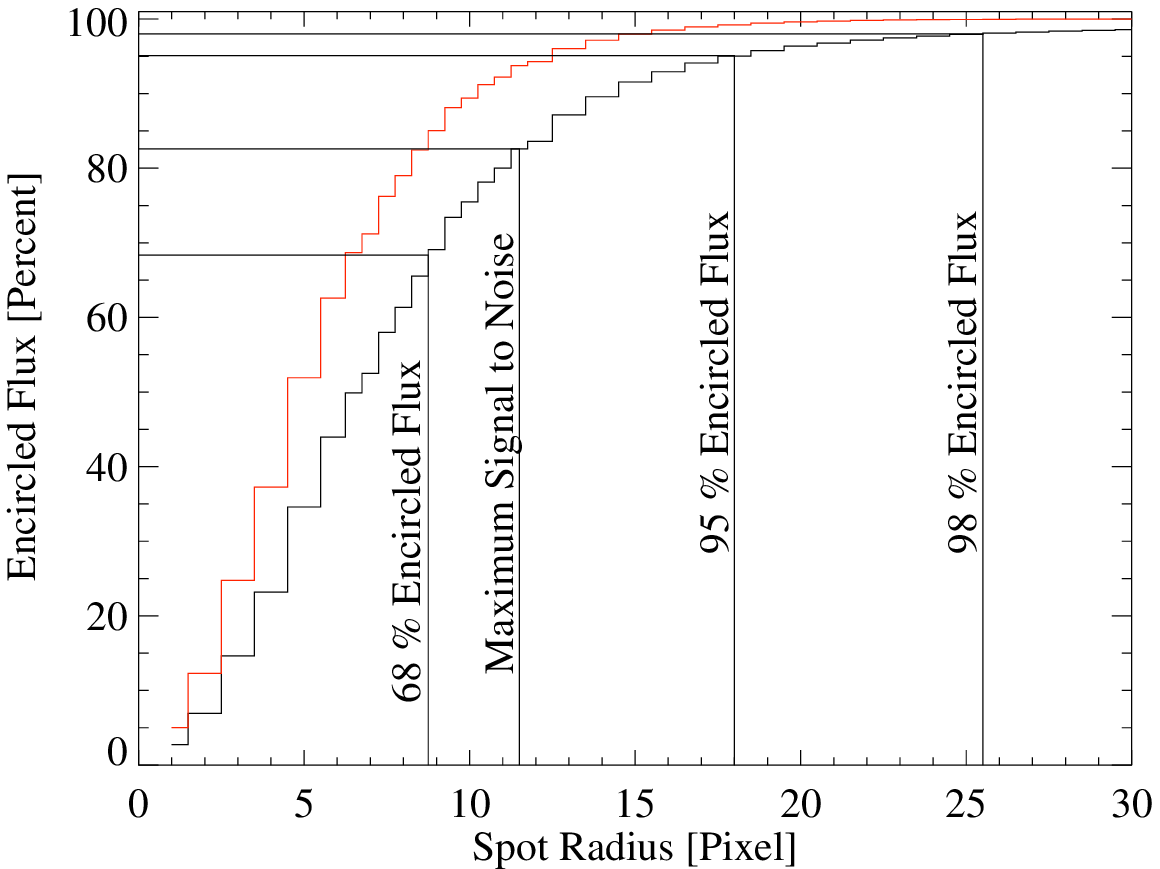}}
  \end{minipage}
  \caption{\label{fig:encircled-intensity} Left: Signal-to-noise ratio for
    the X-ray telescope depending on the radius of the signal-spot. 
           The axion-photon couplings vary from 
           $g_{\text{a}\gamma}=2\times 10^{-10}$~GeV$^{-1}$ for 
           the lowest curve, through 
           $g_{\text{a}\gamma}=5, 9$ to $16\times 10^{-10}$~GeV$^{-1}$ 
           for the curve
           with the largest peak value. 
     Right: The axion flux
    expected in the signal-spot area on the CCD relative to the axion flux
    for an spot radius of the size of the solar disk (encircled flux). Two
    cases are shown: the encircled flux for a perfect linear optics (red
    line) and the encircled flux for a realistic X-ray optics taking into
    account the point spread function of the CAST X-ray mirror system
    (black line).}
\end{figure*}

\subsubsection{Size of the signal extraction region.}
\label{ssub:CCDdata}

\begin{figure*}
  \centering
  \begin{minipage}{0.49\textwidth}
    \centerline{\includegraphics[height=0.6\textheight,angle=0]
      {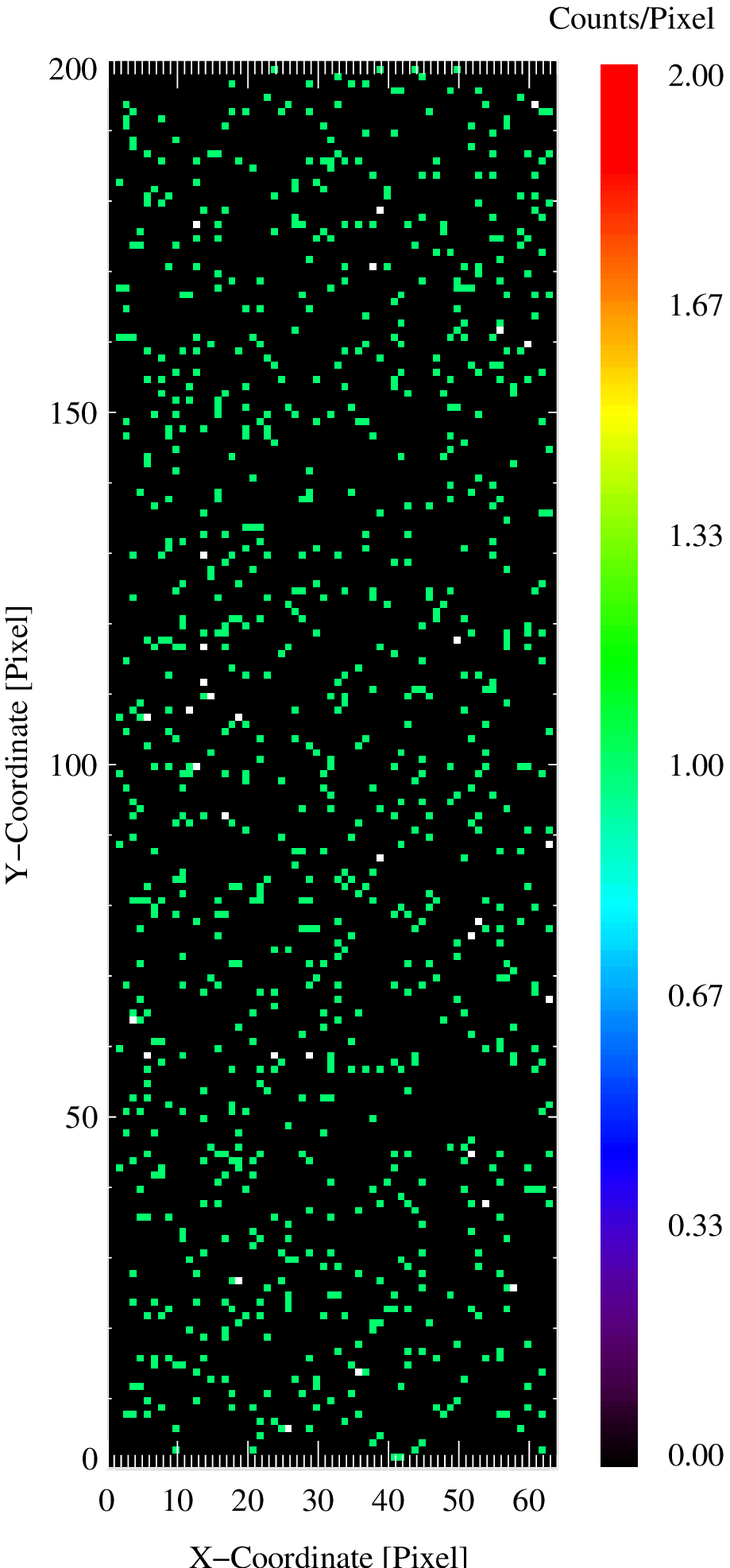}}
  \end{minipage}
  \begin{minipage}{0.49\textwidth}
    \centerline{\includegraphics[height=0.6\textheight,angle=0]{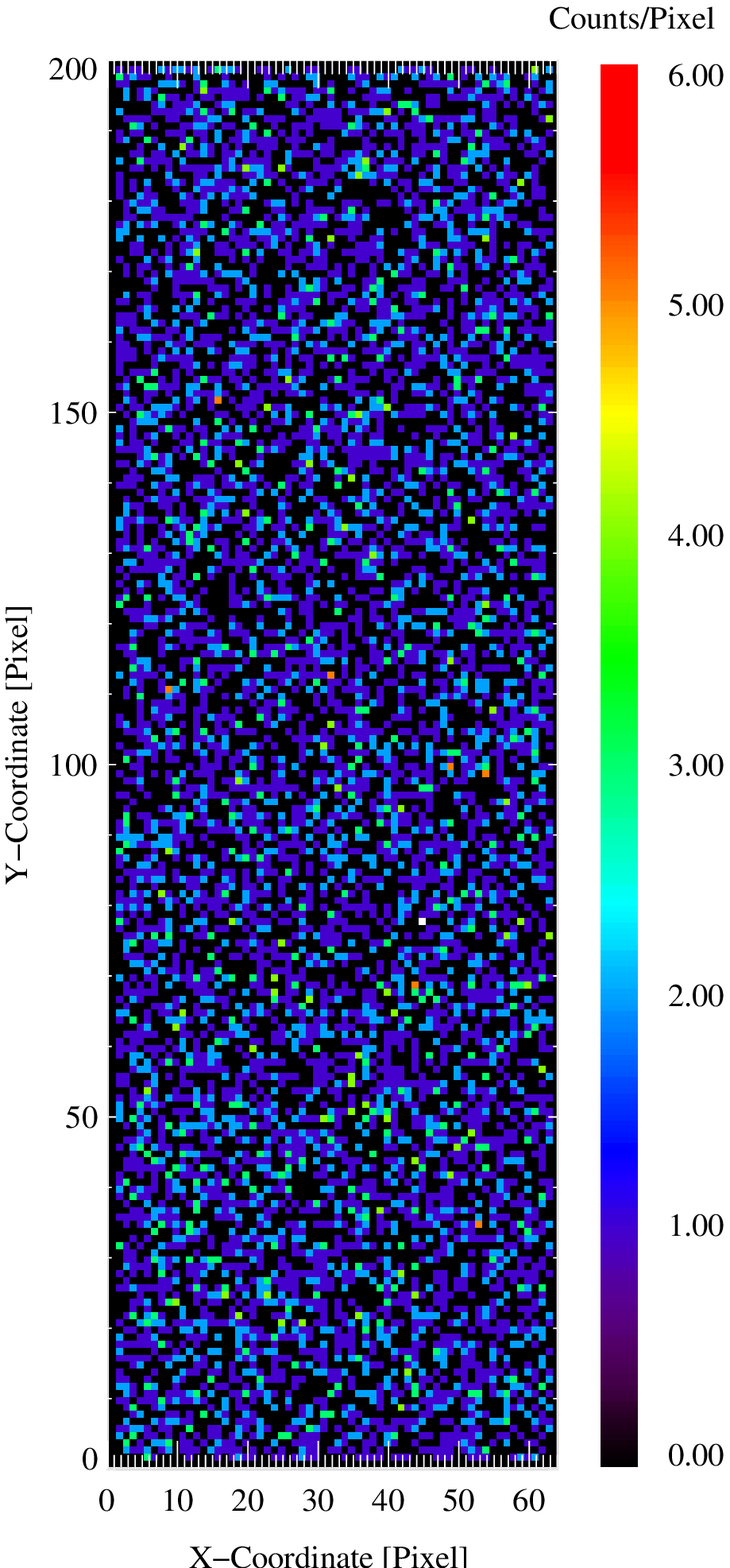}}
  \end{minipage}
  \caption{\label{fig:ccd-background-distribution} Left: Spatial
    distribution of events observed under axion sensitive conditions by the
    CAST X-ray telescope during the 2004 data taking period. The intensity
    is given in counts per pixel and is integrated over the tracking period
    of 197~h. Right: Background spatial distribution as observed by the
    CAST X-ray telescope during the 2004 data taking period. The intensity
    is given in counts per pixel and integrated over the full observation
    period of $1890\,\text{h}$.}
\end{figure*}

For a fixed location of the signal-spot on the CCD we expect that an
optimum spot radius $r_{\text{s}}$ exists, such that the signal-to-noise
ratio of the CCD detector has a maximum. In order to find the best radius,
we used the signal-to-noise ratio depending on $r_{\text{s}}$ for
different \gagg according to
\begin{equation}
  \frac{S}{N}(r_{\text{s}})=\frac{s(r_{\text{s}},\gagg)}
                            {\sqrt{s(r_{\text{s}},\gagg)+b(r_{\text{s}})}}\;,
\end{equation}
where $s(r_{\text{s}},\gagg)$ is the number of counts from axion-to-photon
conversion expected in the signal region for a specific \gagg, calculated
from equation~(\ref{eq:count-spectrum}) by summing over all energy bins in
the $1$--$7\,\text{keV}$ energy range. Further, $b(r_{\text{s}})$ is the
number of background counts we expect in a spot with the same area and
location. To determine $b(r_{\text{s}})$ we used the background data which
provides enough statistics [$b(r_{\text{s}})\gg 200$ counts] to avoid the
optimization to be biased by low counting statistics. The background data
was observed under the same axion sensitive conditions as during tracking
while we were not pointing to the Sun (see right panel of
figure~\ref{fig:ccd-background-distribution}). The left panel of
figure~\ref{fig:encircled-intensity} shows the resulting signal-to-noise
  ratio depending on $r_{\text{s}}$ for different values of \gagg.  We find
  a common maximum for different \gagg at a spot radius of $r_{\rm
    s}=11.5\,\text{pixels}$ which corresponds to 82.6\% encircled solar
  axion flux (see right panel of figure~\ref{fig:encircled-intensity}).

\subsubsection{Spectral fitting results.}
In order to minimize the influence of the Cu-K fluorescence line at
$\approx 8\,\text{keV}$ apparent in the background spectrum of the CCD on
the sensitivity of the X-ray telescope (originating in materials close to
the CCD chip \cite{Kus06}), we restricted our analysis to the energy range
from $1$ to $7\,{\rm keV}$.  In total we observed $26\,\text{counts}$ in
this energy range inside the signal-spot during axion sensitive conditions
(exposure time $197\,\text{h}$). The background is defined by the data
taken from the same signal-spot area during the non-tracking periods,
acquired under the same operating conditions except that the magnet was not
pointing to the Sun. The spatial distribution of the observed events during
tracking and non-tracking time is shown in
figure~\ref{fig:ccd-background-distribution}.

The resulting low counting statistics required the use of a likelihood
function in the minimization procedure to determine an upper limit on
\gagg, rather than a $\chi^2$-analysis. The likelihood function we used is
based on a Poissonian p.d.f. and is given by
\begin{eqnarray}
  \label{eq:ccd-likelihood}
  \cal L(\mu)& = {\prod_i^{N} {\mathrm e}^{-\mu_i}
  \frac{\mu_i^{n_i}}{n_i!}}
  \,\Bigg/\,
  {\prod_i^{N} {\mathrm e}^{-n_i}\frac{n_i^{n_i}}{n_i!}}\;,
\end{eqnarray}
where $N=20$ is the number of spectral bins, $n_i$ the number of observed
counts in bin $i$, and $\mu_{i}$ the value of the fit function in bin $i$.
As a fit function $\mu_{i}=s_{i}+b_{i}$ was used, where $b_{i}$ is the
measured background and $s_{i}$ the expected number of counts in the energy
bin $i$ from axion-to-photon conversion. The best estimate for
$g_{a\gamma}^4$ is obtained by minimizing $S=-2\ln{\cal L(\mu)}$.  In the
large-sample limit the statistic $S$ is $\chi^2$-distributed~\cite{Yao06}.
In the Poissonian regime, but with $N$ relatively large, this is a
reasonable assumption. The validity of the $\chi^2$ interpretation of $S$
in our particular case, as well as the negligible influence of the
statistical uncertainty of the background on the final result have been
verified with a toy Monte Carlo model.  We compared the result derived with
the maximum-likelihood estimator defined in
equation~(\ref{eq:ccd-likelihood}) with different other maximum-likelihood
techniques based on an unbinned maximum-likelihood estimator, a folded
extended likelihood estimator, a binned likelihood estimator that takes
into account the spectral shape of the observed background, and an extended
likelihood estimator with Poissonian convolution. According to our Monte
Carlo simulations all methods are unbiased and yield identical results. We
stress that in the absence of a signal (background-dominated regime), the
upper limit on $g_{{\rm a}\gamma}$ does not improve when taking the
expected intensity distribution in the axion spot shown in
figure~\ref{fig:ccd-backrgound-distribution-axion-spot} into account. This
holds as long as the effective size of the signal region is kept constant.

\begin{figure*}[!ht]
  \begin{center}
  \begin{minipage}{0.49\textwidth}
    \centerline{\includegraphics[width=0.95\columnwidth,angle=0]{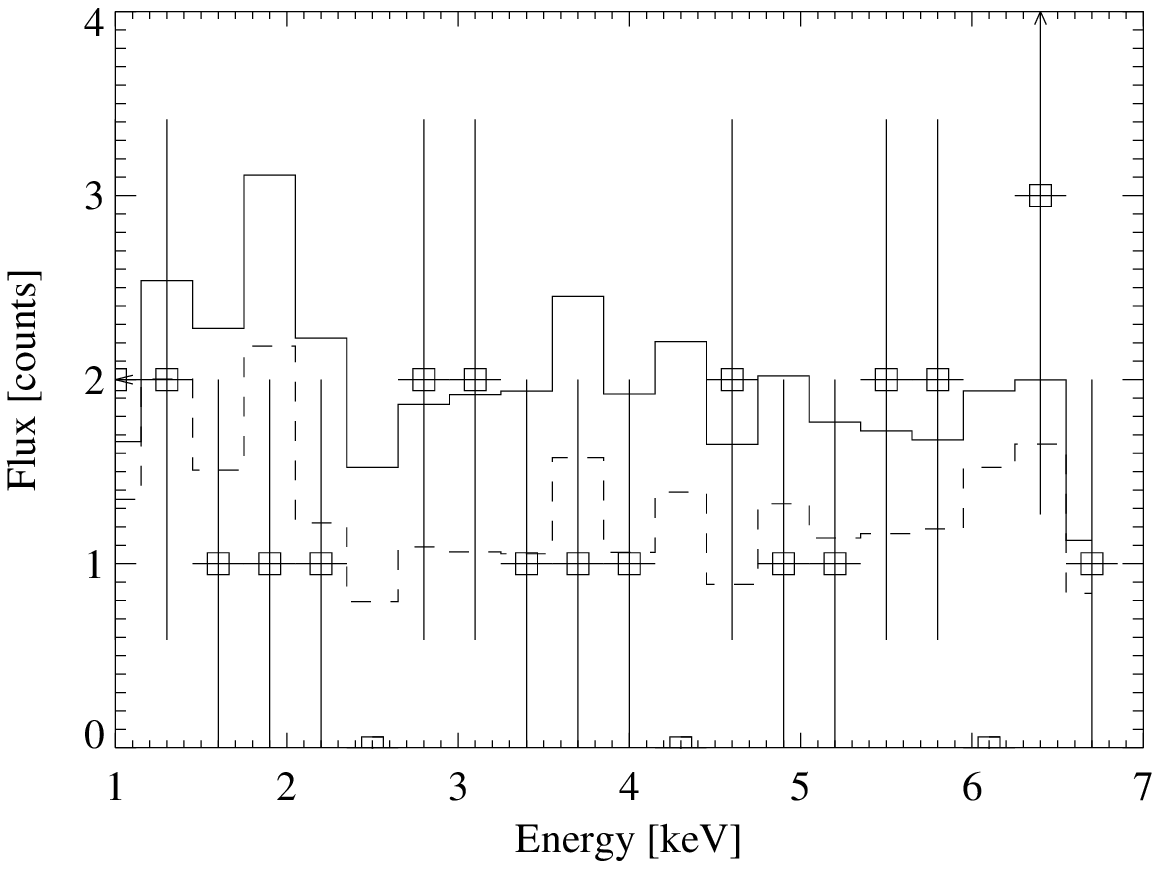}} 
  \end{minipage}
  \begin{minipage}{0.49\textwidth}
    \centerline{\includegraphics[width=0.95\columnwidth,angle=0]{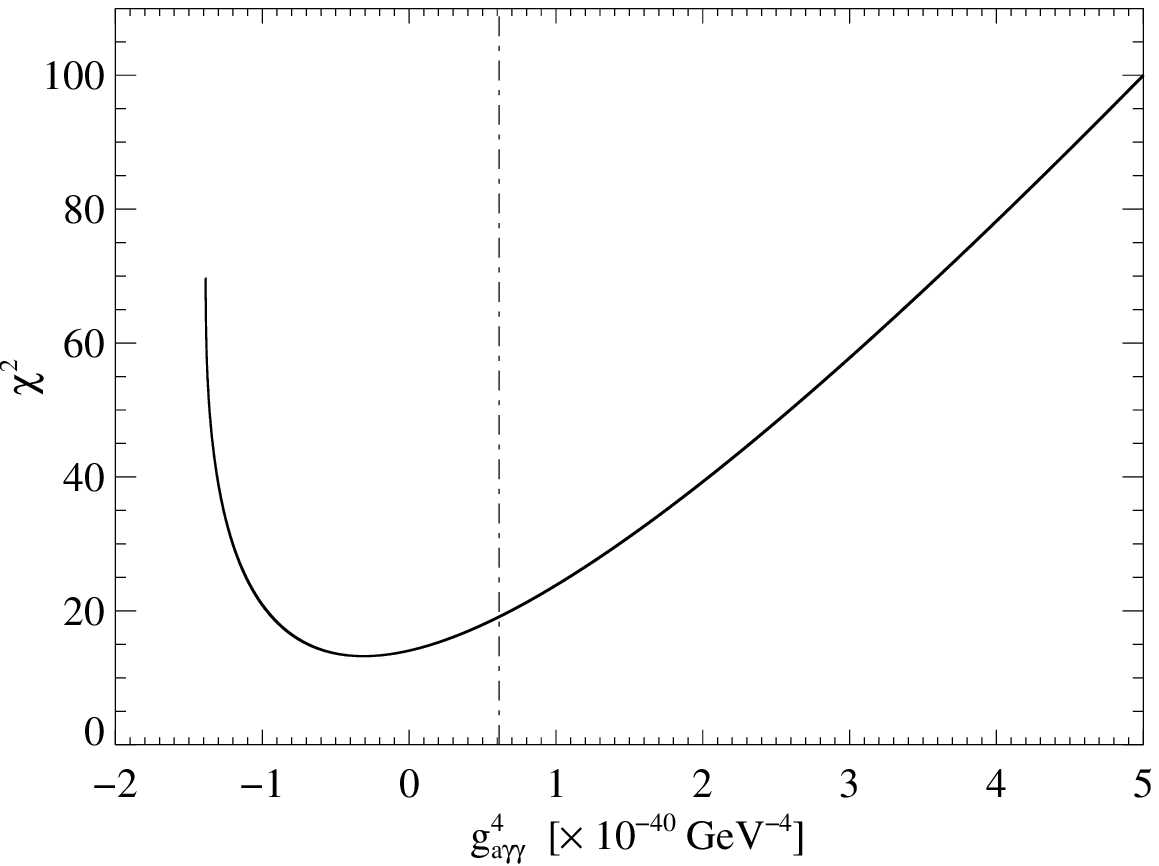}} 
  \end{minipage}
  \end{center}
  \caption{\label{fig:spectral-fit} Left panel: Spectral distribution
    ($\opensquare$) of the data measured with the X-ray telescope during
    the tracking exposure of $197\,\text{h}$ together with the
    expectation $\mu_i$ for the best fit $g_{{\rm a}\gamma}$ ($\broken$)
    and for the 95\% CL limit on $g_{{\rm a}\gamma}$ ($\full$) in units of
    total counts per energy bin in the signal-spot area
    ($9.35\,\text{mm}^2$). Right panel: Corresponding $\chi^2$ distribution
    depending on $g_{{\rm a}\gamma}^4$. The vertical line ($\chain$) marks
    the upper limit on $g_{{\rm a}\gamma}^4 (95\%\,{\rm CL})$.}
\end{figure*}

The final analysis yields an upper limit on the axion-photon coupling of
$g_{{\rm a}\gamma}~<~8.9~\times~10^{-11}\,{\rm GeV}^{-1}$ (95\% CL). The
observed energy spectrum together with the theoretically expected axion
spectrum for the best fit value of $g_{{\rm a}\gamma}$ and for the 95\% CL
limit on $g_{{\rm a}\gamma}$ is shown in figure~\ref{fig:spectral-fit}
together with the likelihood distribution of the fit. A summary of our
results is given in table~\ref{data04}.

\subsubsection{Systematic uncertainties.}
We studied the influence of systematic uncertainties on the best fit value
of $g_{{\rm a}\gamma}^4$ and on the upper limit on $g_{{\rm a}\gamma}$. As
stated in \cite{Kus06} statistically significant variations of the
background on long (months) and short time scales (hours to days) are not
apparent from the X-ray telescope data. Even if the background level would
be time-dependent, it would not play a significant role, since we
measure the potential signal (signal-spot) and background (area around the
signal-spot) simultaneously. A crucial point in the analysis is the
definition of the area where the background is taken from. Ideally, if the
background is homogeneous spatially distributed, the area surrounding and
close to the signal spot could be used. Since we observed no significant
difference between the background level and spatial distribution during
tracking and non-tracking times, we used the same signal-spot area during
the non-tracking periods to define the background.  Alternatively, we used
tracking data and different regions on the CCD outside the signal-spot area
to estimate the systematic uncertainties due to the choice of the
background definition.  
The overall systematic uncertainty is dominated by 
both the background definition and the pointing accuracy, that affects
the effective area of the telescope, the location of the signal-spot and its
size respectively. Other effects such as uncertainties in
the detector calibration and magnet parameters are negligible.  
For the best fit value
of $g_{{\rm a}\gamma}^4$ we find
\begin{equation*}
  (g_{{\rm a}\gamma}^4)_{\rm best fit} =
  (-0.31{\pm^{0.35}_{0.29}}_{\text{stat}}
   \pm0.28_{\text{syst}})\times 10^{-40}\,\text{GeV}^{-4}\,. 
\end{equation*} 
 An effect of the systematic uncertainties on the upper limit 
 on $g_{{\rm a}\gamma}$ is estimated to be less than $\sim$ 5\%. 

\subsection{TPC and MM detector}               \label{sub:TPCmm}

The TPC~\cite{Aut07}, covering both bores of the east 
end of the magnet and facing ``sunset'' axions, was
operated continuously during the entire 2004 data taking period and
acquired $203\,{\rm h}$ of tracking data and $2616\,{\rm h}$ of 
background data.  The main
difference between the 2003 and 2004 experimental setup were the operation
of a new differential pumping system and an improved passive shield for the
detector. The purpose of the differential pumping system is to decrease the
effect of gas leaks towards the magnet and to minimize the possibility of
damage to the TPC windows due to sudden pressure changes or break down. As
a result of the smooth operation of the TPC detector, the exposure time
during 2004 was five times higher compared to 2003. The TPC is housed in a
low radioactive, 5~mm thick copper box inside a low radioactive
multicomponent shield (22~cm of polyethylene, 1~mm of Cd and 2.5~cm of Pb).
The entire shield is permanently flushed with nitrogen, creating an over
pressure which decreases the radon contamination close to the detector.
The installation of the shield has reduced the TPC background level in the
region between 1 and 10 keV by a factor of 4.3 compared to the background
level of 2003. The typical differential counting rate in this energy
interval is $4.1 \times 10^{-5}$ counts cm$^{-2}$ s$^{-1}$ keV$^{-1}$. One
of the most important advantages of the improved background suppression is
that the observed background is almost independent of the magnet pointing
direction, which was not the case during the 2003 data taking period.  

The micromegas detector is placed behind the other bore at the
west end of the magnet. The new MM~\cite{Abb07} was specifically
designed to eliminate the cross talk effects present at the previous
detector and to improve the quality of the strips. The smooth
operation of the improved MM detector during the 2004 data taking
period, combined with the development of more effective off-line
analysis techniques, allowed for further reduction of the background
level by a factor of 2.5 compared to the 2003 running period. Due to
the stability of the detector and the experiment in general, it was
possible to accumulate $196\,{\rm h}$  of tracking data and 
$3000\,{\rm h}$ of
background data. The resulting background counting rate is $5\times
10^{-5}$~counts~cm$^{-2}$~s$^{-1}$~keV$^{-1}$ in the 1--8.5~keV
energy range.
\begin{table*}[t]
  \centering
  \footnotesize
\caption{\label{data04} 2004 and 2003~\cite{Zio05}
   data sets included in our result.}
  \begin{tabular}{cccccccc}
  \br 
    \multicolumn{1}{c}{Run} & \multicolumn{1}{c}{Detector}  
    & \multicolumn{1}{c}{Tracking} &
    \multicolumn{1}{c}{Background} &
    \multicolumn{1}{c}{($g^4_{{\rm a}\gamma})_{\rm best\, fit}
         \pm 1\sigma_{\rm stat}$} &
    \multicolumn{1}{c}{$\chi^2_{\rm min}$/d.o.f} &
    \multicolumn{1}{c}{$g_{{\rm a}\gamma}$ (95\% CL)} \\ 
    & & (h) &
    (h) & $(10^{-40}$ GeV$^{-4})$ & & $(10^{-11}$ GeV$^{-1})$\\ 
 \mr
    2004 & CCD & 197   & 1890   & $-0.31\pm^{0.35}_{0.29}$ & 13.2/19 & 8.9\\
         & TPC & 203   & 2616   & $1.04\pm^{1.02}_{1.01}$ & 17.6/17  & 12.9\\
         & MM  & 196   & 3000   & $0.21\pm^{1.27}_{1.26}$ & 24.7/14 & 12.7\\
 \mr
    2003 & CCD & 121.3 & 1233.5 & $0.4\pm1.0$ & 28.5/19 & 12.3\\
         & TPC &  62.7 &  719.9 & $-1.1\pm3.3$ & 18.1/17 & 15.5\\
         & MM set A & 43.8 & 431.4 & $-1.4\pm4.5$ & 12.4/13 & 16.7\\
         & MM set B & 11.5 & 121 & $2.5\pm8.8$& 6.1/13 & 20.9\\
         & MM set C & 21.8 & 251 & $-9.4\pm6.5$& 10.7/13 & 16.7\\ 
 \br
  \end{tabular}
\end{table*}

The information of the energy scale for each detector and each run
configuration is assured by periodical calibrations with the
$^{55}$Fe X-ray source.  The hardware efficiency with respect to the
photon energy has been simulated using the GEANT4 toolkit and
verified by theoretical calculations and experimental data from the
detectors that were characterized in the PANTER test facility in
Munich~\cite{Freyberg:06a}.

\subsubsection{TPC and MM data analysis and results.}
\label{ssub:TPC-MMdata}
\begin{figure*}[!ht]
  \begin{minipage}{0.49\textwidth}
    \centerline{\includegraphics[width=0.95\textwidth]{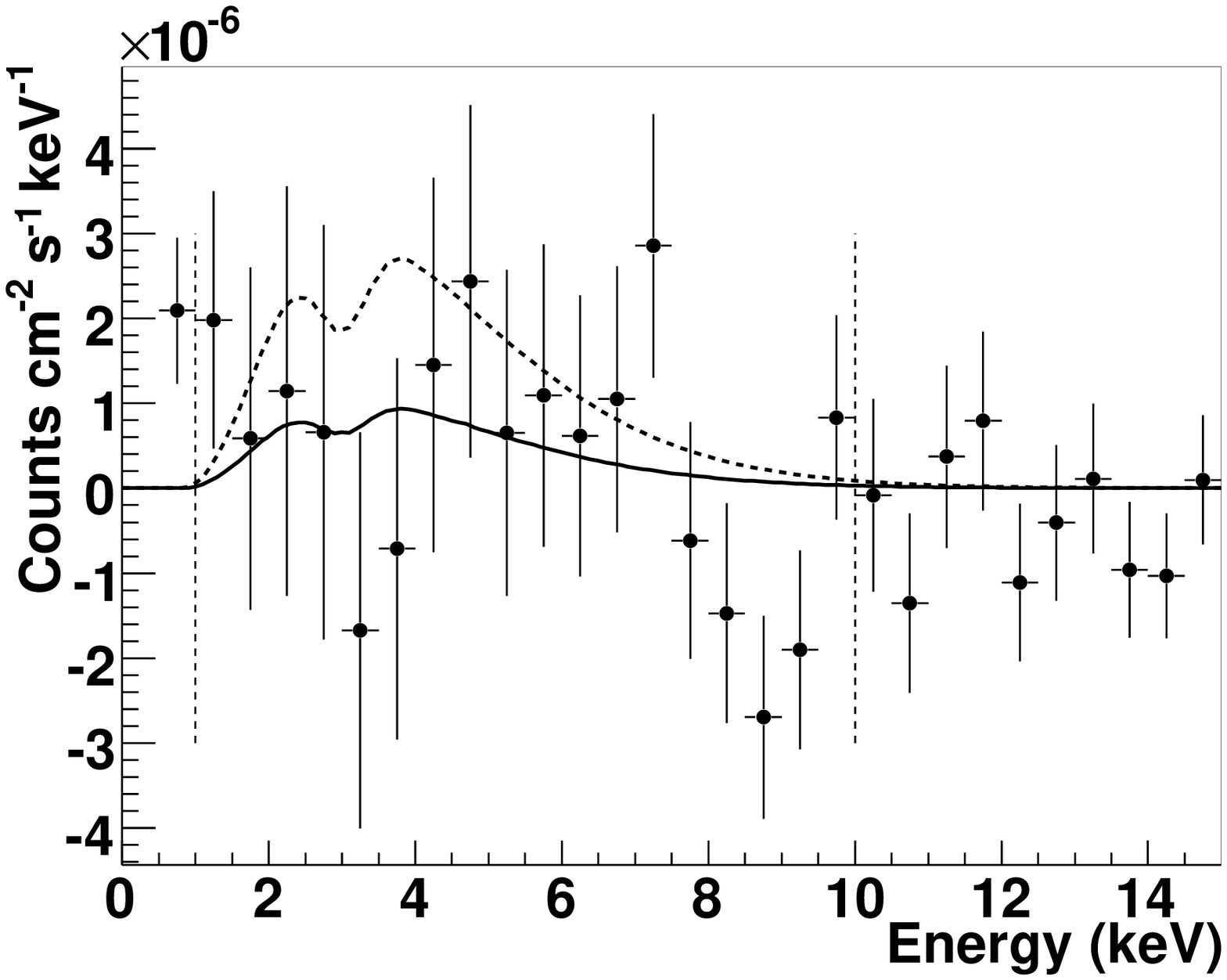}}
  \end{minipage}
  \begin{minipage}{0.49\textwidth}
    \centerline{\includegraphics[width=0.95\textwidth]{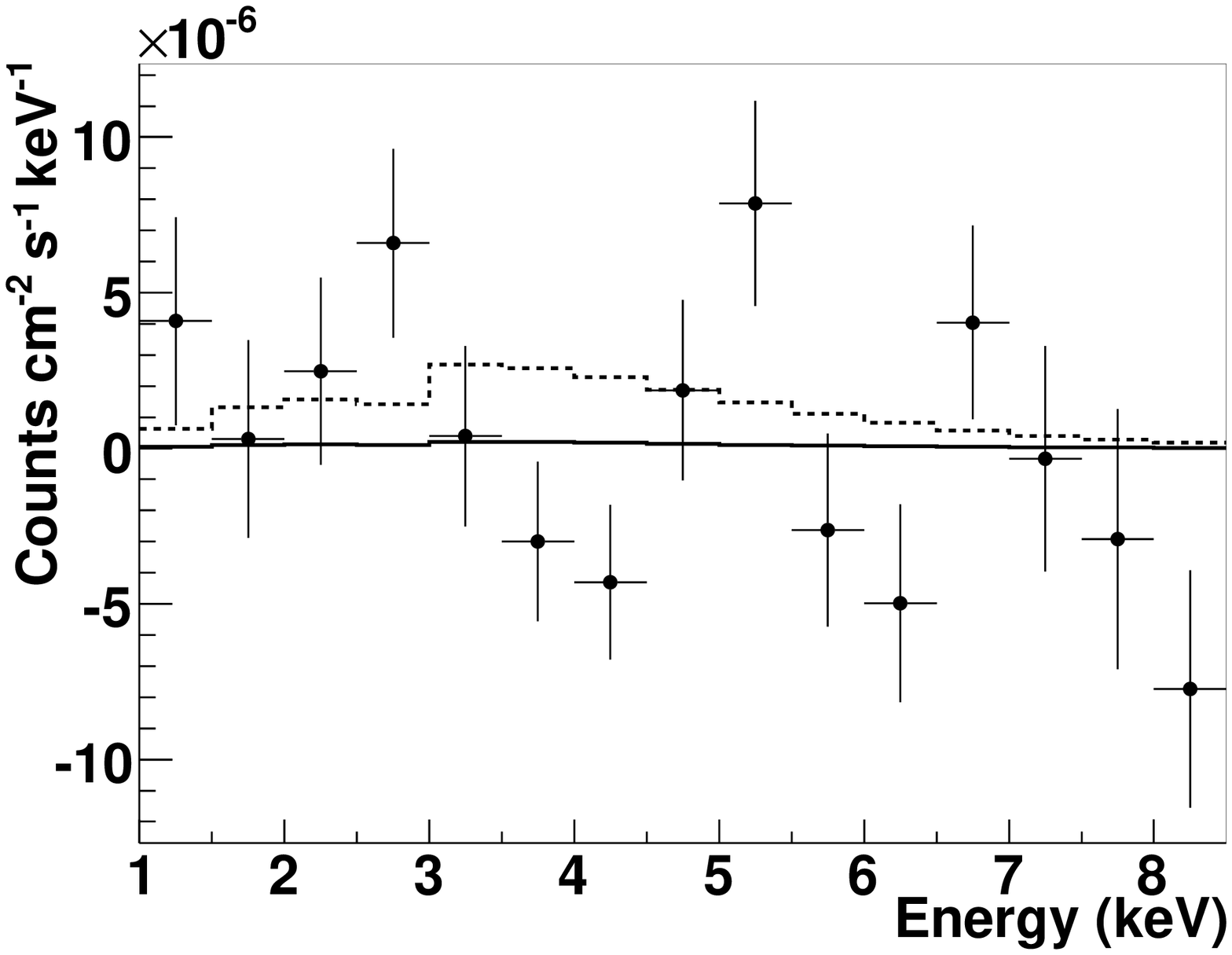}}
  \end{minipage}
 \caption{
   Panels (left) and (right) show the experimental subtracted spectrum
   ($\fullcircle$) together with the expectation for the best fit $g_{{\rm
       a}\gamma}$ ($\full$) and for the 95\% CL limit on $g_{{\rm
       a}\gamma}$ ($\dashed$) for the TPC data set and MM data set,
   respectively.  The structure at 3~keV in the expected spectrum reflects
   the change in the efficiency curves of the gaseous detectors due to the
   Ar K-edge.  The vertical dashed lines indicate the fitting window. The
   fully coherent conversion regime ($m_{\rm a} \lesssim 0.02$~eV) is
   assumed.
 \label{tpc}}
\end{figure*}

The results presented here were obtained after the analysis of the
TPC and MM 2004 data sets listed in table~\ref{data04}.
The analysis was performed by standard $\chi^2$ minimization.
The accumulated effective background spectrum, properly normalized, was
subtracted from the corresponding tracking spectrum and the resulting data
was fitted with the theoretical axion signal\footnote{The expected spectrum
of axion-induced photons was calculated by equation~(\ref{eq5}) and
multiplied by the detection efficiency curves.}
which scales with $g_{{\rm a}\gamma}^4$ for various $m_{\rm a}$.
The obtained results are given in table~\ref{data04}, showing no significant
excess of axion-to-photon conversion-like events. Therefore, an upper limit on
$g_{{\rm a}\gamma}$ at
95\% confidence level was calculated for each of the data sets following
the Bayesian scheme (shown in the last column of table~\ref{data04}).
The corresponding fits are plotted in figure~\ref{tpc}.

The experimental systematic uncertainties on the present limits
have been studied. The daily
variation of the background caused by the environmental radon is the
dominant factor in the study of systematics
for the MM analysis. Most of its effect is eliminated by using background data
taken at times of the day as close as possible as the tracking times. The
resulting upper limits for different choices of effective backgrounds clearly
suggest that the systematic uncertainty is less than 2\%. Systematic effects
in the fitting procedure of the TPC data are considered by varying the
background level until the null hypothesis test (in areas of the
detector where no signal is expected) yields a result
with a probability smaller than 5\%. If taken as an uncertainty, this range
corresponds to $\sim$ 10\% variation of the upper limit.

\section{Combined result of CAST phase I data sets}
\label{sec:CASTphaseI}

\begin{figure*}[t]
\begin{center}
 \centerline{\includegraphics[width=0.65\textwidth]{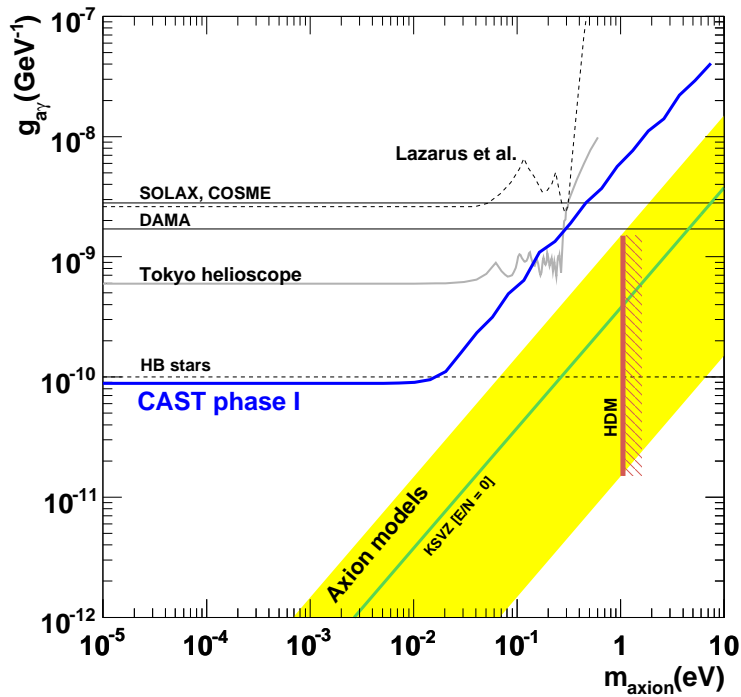}}
 \caption{ 
   Exclusion plots in the axion-photon coupling versus the axion mass
   plane. The limit achieved by the present experiment (CAST phase~I) is
   compared with other constraints (Lazarus et al.~\cite{Laz92}, 
   SOLAX~\cite{Avi98}, COSME~\cite{Mor02}, DAMA~\cite{Ber01}, 
   Tokyo helioscope~\cite{Mori98,Ino02} and HB 
   stars~\cite{Raffelt:2006cw,Raf96}) 
   discussed in the introduction. The vertical red line (HDM) is the
   hot dark matter limit for hadronic axions 
   $m_{\rm a}<1.05~{\rm eV}$~\cite{Han05} inferred
   from observations of the cosmological large-scale structure.
   The yellow band represents typical theoretical models with 
   $\left|E/N-1.95\right|$ in the range 0.07--7 while the green solid line
   corresponds to the case when $E/N=0$ is assumed.  
 \label{exclusion}}
 \end{center}
\end{figure*}

The best fit values of $g_{{\rm a}\gamma}^4$ obtained for each of
the detector's 
2004 data are reported in table~\ref{data04} together with
their $1\sigma$ error and the corresponding $\chi^2_{\rm min}$
values and degrees of freedom. Each of the data set is individually
compatible with the absence of any signal. A combined result has
been obtained by multiplying the Bayesian probabilities of the three
fits to get a combined probability function and obtaining the value
of $g_{{\rm a}\gamma}^4$ that encompasses 95\% of its physically
allowed (i.e. positive signals) part. The result of this operation is
$g_{{\rm a}\gamma} < 9.0\times 10^{-11}~{\rm GeV}^{-1}$ at 95\% CL.
The same operation, including in the multiplication also the Bayesian
probability functions of the 2003 data sets~\cite{Zio05} listed 
in table~\ref{data04}, gives a final limit for the CAST vacuum setup of 
\begin{equation}\label{finalresult}
 g_{{\rm a}\gamma} < 8.8
           \times 10^{-11}~{\rm GeV}^{-1}\quad(95\%\,{\rm CL})\,. 
\end{equation}
This value is valid for $m_{\rm a}\lesssim 0.02$~eV where the expected
signal is mass independent because the axion-photon oscillation length far
exceeds the length of the magnet. For higher $m_{\rm a}$, the overall
signal strength diminishes rapidly and the spectral shape differs (see
equation~(\ref{eq4})). The described procedure
was repeated for different values of $m_{\rm a}$ to get the entire 95\% CL
exclusion line.  The region excluded in the $g_{{\rm a}\gamma}$--$m_{\rm a}$ 
plane is shown in figure~\ref{exclusion} together with the
results of other experiments as well as astrophysical 
and cosmological
constraints. The statistical limit on the sensitivity is mostly set 
by the value of $(BL)^2$ in the LHC test magnet.  
In general, the systematic uncertainties on the background 
spectrum due to its position and time dependence and due to the pointing
accuracy are estimated to have an effect of less than $\sim$~5\% in the 
final upper limit obtained, that is dominated by the result of the X-ray 
telescope.  

\section{Conclusion}                              \label{sec:conclusion}

We have searched for the coherent production of photons in a
decommissioned LHC test magnet of $L=9.26$ m and $B=9$ T, assuming
the existence of an incident axion flux that is emitted from the Sun
by the Primakoff process. Improvements to this experiment relative
to the published 2003 results come from several changes in the
detectors, in the shielding, and in the operating conditions.

In conclusion, we obtained the currently best experimental limit on
the axion-photon coupling which is by a factor 1.3 more
restrictive than the CAST limit previously reported~\cite{Zio05}.
For axion masses less than 0.02~eV it supersedes the previous
astrophysical limit based on the helium-burning lifetime of HB
stars, as shown in figure~\ref{exclusion}. Our sensitivity does not
yet touch the band of $g_{{\rm a}\gamma}$--$m_{\rm a}$ values
expected for QCD axion models and thus applies to somewhat
lighter axion-like particles.

Our limits can be relevant, for example, in comparison with the
anomalous signature, far in excess of the QED expectation of vacuum
birefringence, that has been reported by the PVLAS
experiment~\cite{Zav05}. An interpretation of this signal in terms
of axion-like particles requires a coupling strength of about
$10^{-5}$--$10^{-6}~\rm{GeV}^{-1}$ for $m_{\rm a} \sim 10^{-3}$~eV.
However, this coupling is so large that the solar axion production
would lead to the Sun burning out within about a 1000~years.
Therefore, if the particle interpretation of PVLAS is correct,
something about the underlying physics must be very different than
assumed here~\cite{Masso:2005ym,Masso:2006gc,Antoniadis:2006wp,%
Mohapatra:2006pv,Jaeckel:2006xm,Redondo:2006xx}. Without a specific
underlying theoretical model it is impossible to compare PVLAS and
CAST in generic terms---the discrepancy is simply too large.

In order to extend our search to QCD axions, our setup has
already been upgraded and the data collection has started in late
2005. The CAST experiment, being operated in a scanning mode in
which the He gas pressure is varied in appropriate steps, will allow
us to explore the range of possible axion masses up to about 1~eV.
For the first time an experiment providing a sensitivity
to $g_{{\rm a}\gamma}$ of the order of that derived from HB stars
will be able to enter into the region, shown in
figure~\ref{exclusion}, which is favored by axion models. In this
mass range axions would provide a hot-dark matter component of the
universe, similar to neutrinos~\cite{Mor98,Han05}. Another
interesting feature is that in this regime of operation, when the
gas is inserted in the magnet pipes, CAST could be also sensitive to
the existence of (two) large extra dimensions with the
compactification radius down to around 200~nm.

\ack We thank CERN for hosting the experiment and for the contributions of
J.~P.~Bojon, F.~Cataneo, R.~Campagnolo, G.~Cipolla, F.~Chiusano,
M.~Delattre, F.~Formenti, M.~Genet, J.~N.~Joux, A.~Lippitsch, L.~Musa,
R.~De~Oliveira, A.~Onnela, J.~Pierlot, C.~Rosset, H.~Thiesen 
and B.~Vullierme. We thank in
particular F.~James for his advice concerning the statistical treatment of
the data. We acknowledge support from NSERC (Canada), MSES (Croatia)
under the grant number 098-0982887-2872, CEA
(France), BMBF (Germany) under the grant numbers 05 CC2EEA/9 and 05
CC1RD1/0, the Virtuelles Institut f\"ur Dunkle Materie und Neutrinos --
VIDMAN (Germany), GSRT (Greece), RFFR (Russia), CICyT (Spain), NSF (USA),
US Department of Energy, NASA under the grant number NAG5-10842
and the helpful discussions within the network on direct dark matter
detection of the ILIAS integrating activity (Contract number:
RII3-CT-2003-506222).

\section*{References} 

\end{document}